%% file: main.tex
\documentclass[sigconf,twocolumn]{acmart}   	

\usepackage{geometry}                		
\usepackage{booktabs}   

\usepackage{pgffor}
 \usepackage{xcolor}
\usepackage{url}
\usepackage{comment}
\usepackage{stmaryrd}
\usepackage{amsmath}
\usepackage{amsthm}
\usepackage{mathtools}
\usepackage{amstext}
\usepackage{mathpartir}
\usepackage{listings}
\usepackage{color}
\usepackage{framed}
\usepackage{tabularx}
\usepackage{graphicx}				
\graphicspath{{./figures/}}

\usepackage{enumitem}

\input{macros.tex}

\input{ces-macros.tex}
\input{figure_macros.tex}

\renewcommand{\paragraph}[1]{\vspace*{3pt}\noindent\textbf{#1}}
\usepackage{amsmath}
\usepackage[ruled, noend, linesnumbered]{algorithm2e}
\usepackage{amsthm}

\newtheorem{property}{Property}

\title{Secret Sharing Sharing For Highly Scalable Secure Aggregation}
\author{Timothy Stevens}
\affiliation{University of Vermont}
\email{timothy.stevens@uvm.edu}
\author{Joseph Near}
\affiliation{University of Vermont}
\email{jnear@uvm.edu}
\author{Christian Skalka}
\affiliation{University of Vermont}
\email{christian.skalka@uvm.edu}

\begin{document}
\title{Secret Sharing Sharing For Highly Scalable Secure Aggregation}
\begin{abstract}
    Secure Multiparty Computation (MPC) can improve the security and privacy of data owners while allowing analysts to perform high quality analytics. \emph{Secure aggregation} is a secure distributed mechanism to support federated deep learning without the need for trusted third parties. In this paper we present a highly performant secure aggregation protocol with sub-linear communication complexity for application in large federations.

    Our protocol achieves greater communication and computation efficiencies through a group-based approach. It is similar to secret sharing protocols extended to vectors of values- aka gradients- but within groups we add an additional layer of secret sharing of shares themselves- aka sharding. This ensures privacy of secret inputs in the standard real/ideal security paradigm, in both semi-honest and malicious settings where the server may collude with the adversary. 

    In the malicious setting with 5\% corrupt clients and 5\% dropouts, our protocol can aggregate over a federation with 100,000,000 members and vectors of length 100 while requiring each client to communicate with only 350 other clients. The concrete computation cost for this aggregation is less than half a second for the server and less than 100ms for the client. 
\end{abstract}

%

\keywords{MPC; Differential Privacy; Machine Learning}

\maketitle

\input{intro}

\input{background}

\input{protocol}

\input{probability}

\input{evaluation}

\input{conclusion_new}

\bibliographystyle{ACM-Reference-Format}
\bibliography{biblio}


\end{document}

%% file: macros.tex
\newcommand{\fpkeyword}[1]{\text{\sf #1}}

\newcommand{\stack}[1]{\mathord\mid #1\kern.2mm\mathord\mid}

\newcommand{\cod}[1]{\llbracket#1\rrbracket}

\newcommand{\tset}[1]{\{#1\}}
\newcommand{\tnext}{\,;\;}
\newcommand{\drow}[3]{#1:#2\tnext#3}
\newcommand{\urow}[1]{\partial#1}

\newcommand{\capab}[1]{\mathbf{#1}}

\newcommand{\tpre}{\capab{Pre}}
\newcommand{\tabs}{\capab{Abs}}

\newcommand{\system}[2]{\mathcal{S}_{#1}^{#2}}



%% file: figure_macros.tex
\newcommand\figneighbors{
\begin{figure*}[!htb]
    \makebox[\textwidth][c]{\includegraphics[width=1.2\textwidth]{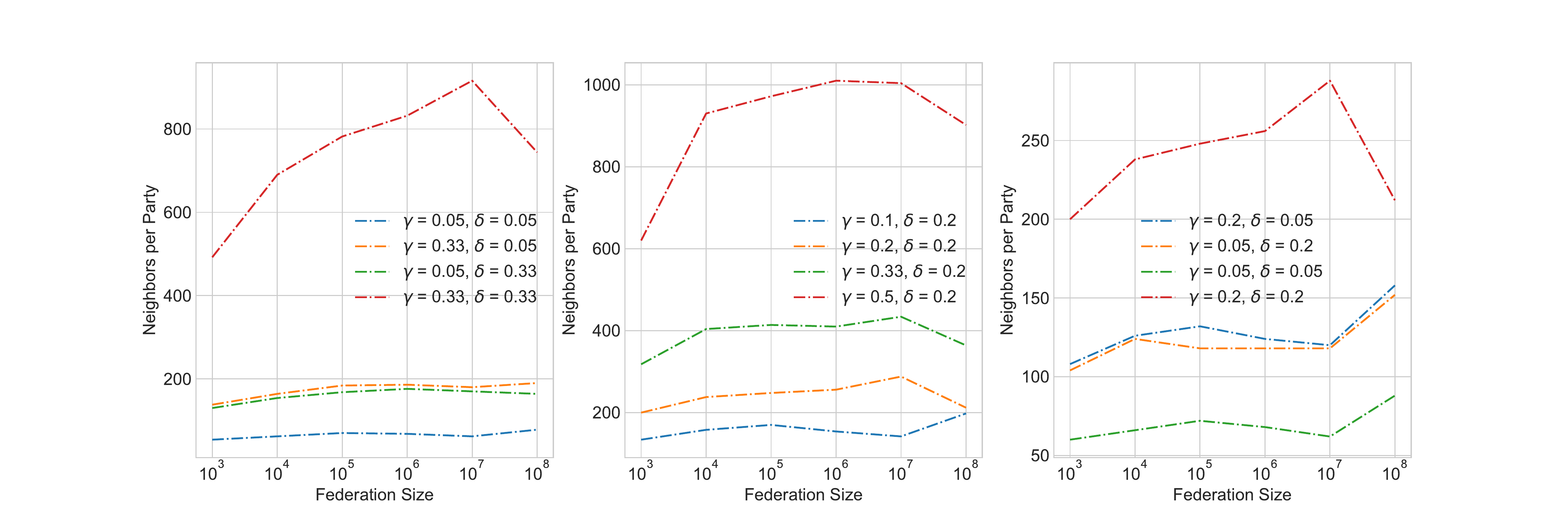}}
    \caption{Number of neighbors required for Protocol~\ref{prot:shard} with different federation sizes and assumptions about the adversary. Left: Semi-honest protocol with less-dropout tolerable configurations ($\eta = -20$, $\delta >= .05$). Center: Semi-honest protocol with more conservative parameter settings. In this case $\eta = -30$, and the highest tested $\gamma$ is $.5$. Right: Malicious protocol configurations.}

    \label{fig:neighbors_dp}
\end{figure*}
}

\newcommand\figbellexpansion{
\begin{figure}[!htb]
    \includegraphics[width=\linewidth]{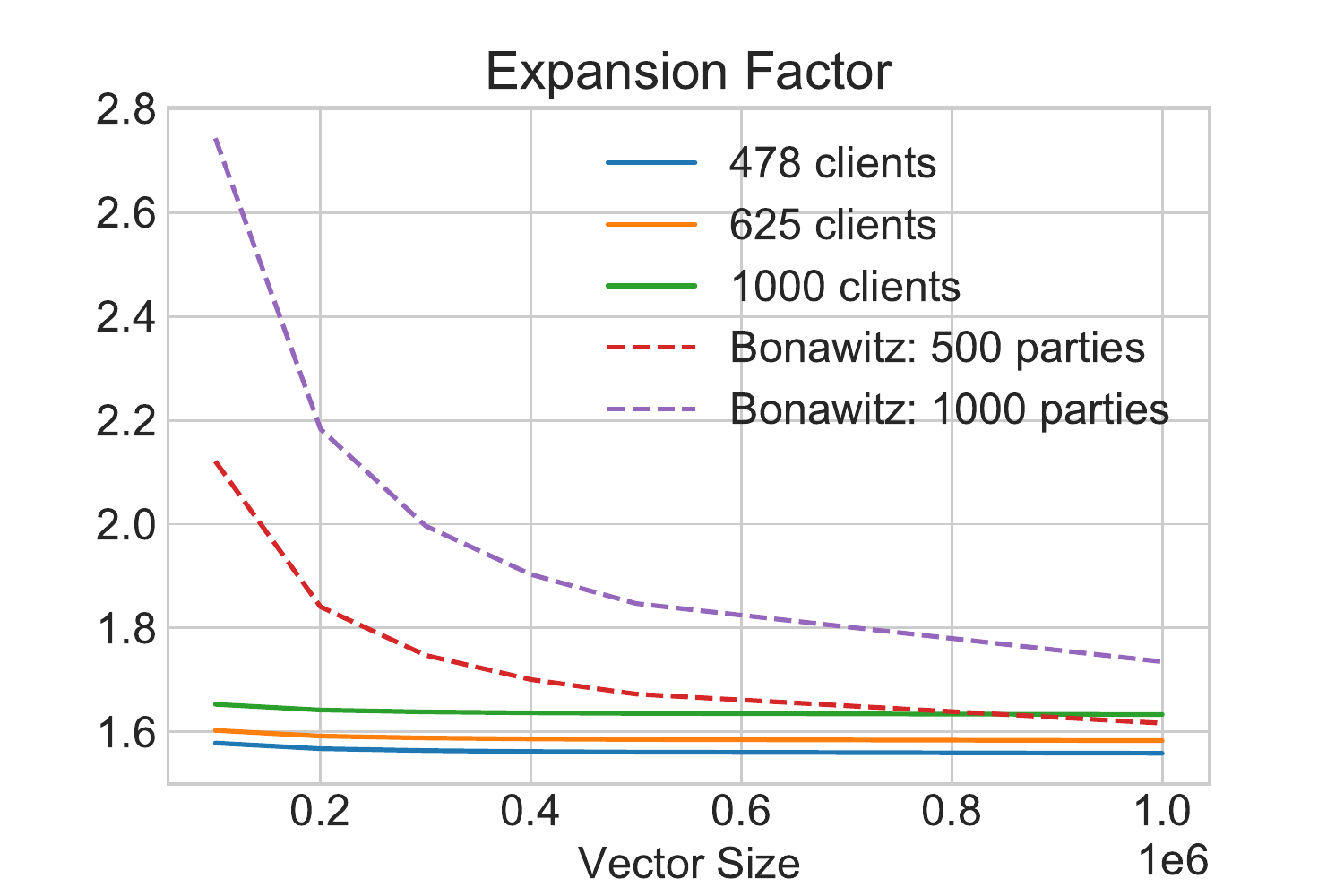}
    \caption{Expansion Factor of protocol defined by~\cite{bell_paper}  and \ourprot with different numbers of neighbors. \ourprot results for larger number of neighbors are reported because \ourprot typically requires about twice the number of neighbors as Bell.}
    \label{fig:expansion_bell}
\end{figure}
}

\newcommand\figexpansion{
\begin{figure*}[!htb]
    \makebox[\textwidth][c]{\includegraphics[width=1.2\textwidth]{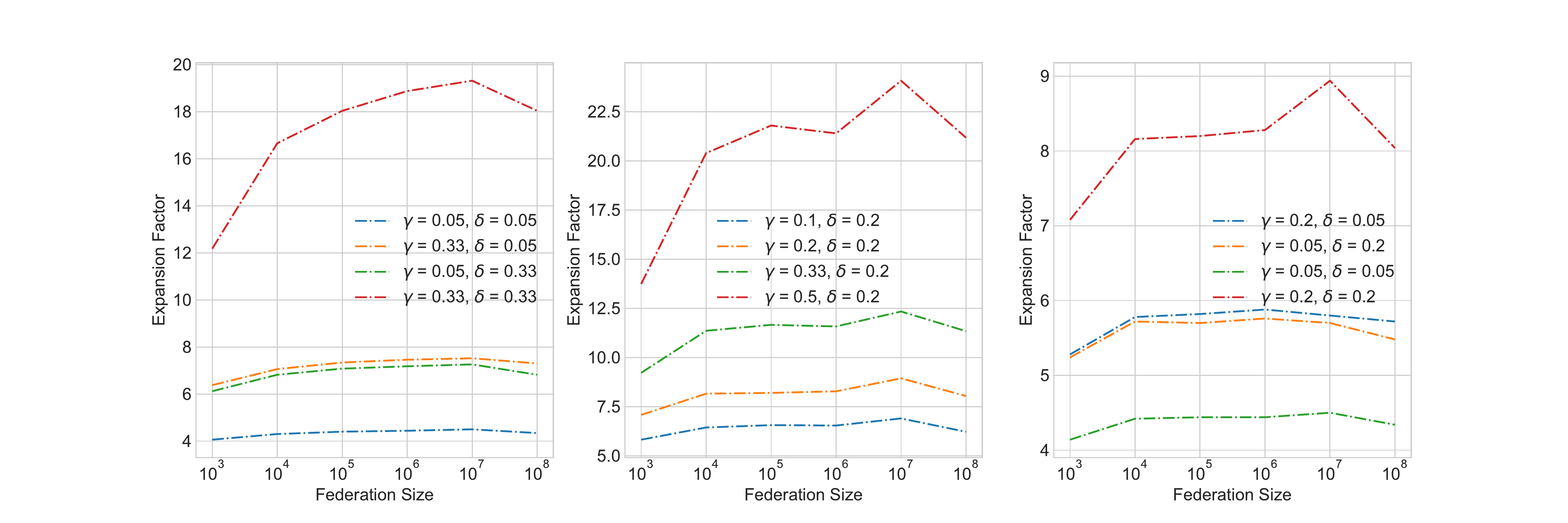}}
    \caption{Expansion Factor of Protocol~\ref{prot:shard} with different federation sizes and assumptions about the adversary. Packed secret sharing is used with $k=100$. Left: Semi-honest protocol with less-dropout tolerable configurations ($\eta = -20$, $\delta >= .05$). Center: Semi-honest protocol with more conservative parameter settings. In this case $\eta = -30$, and the highest tested $\gamma$ is $.5$. Right: Malicious protocol configurations.}
    \label{fig:expansion_dp}
\end{figure*}
}
\newcommand\figserver{
\begin{figure*}[!htb]
    \makebox[\textwidth][c]{\includegraphics[width=1.2\textwidth]{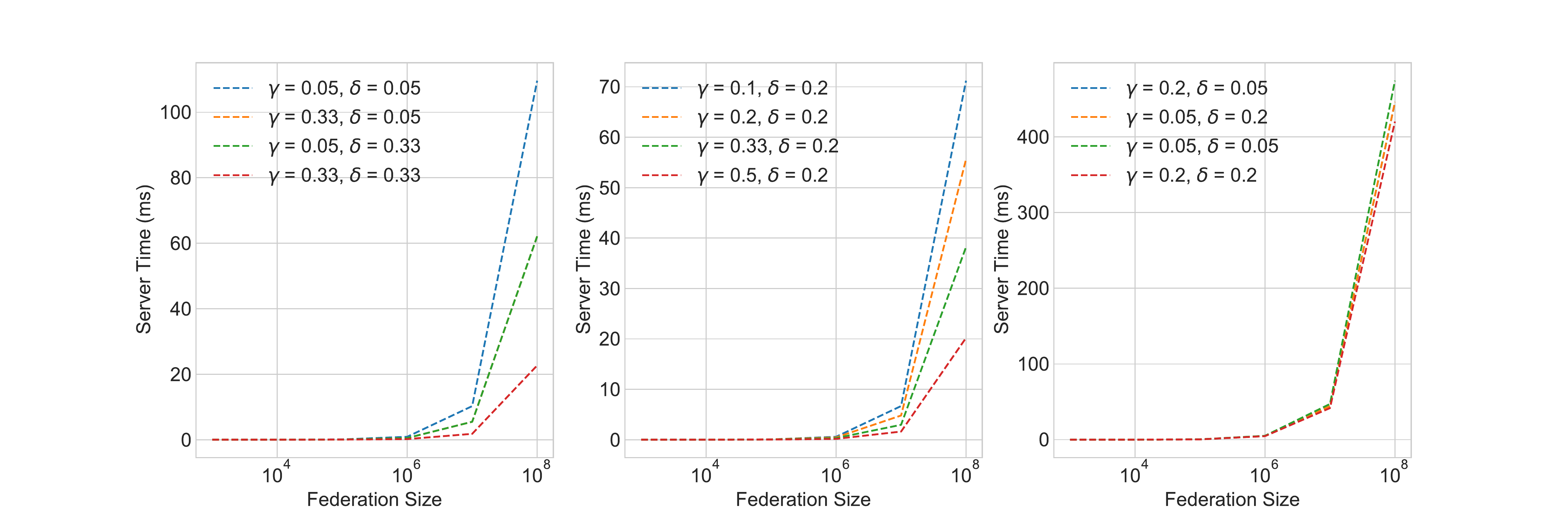}}
    \caption{Server time for Protocol~\ref{prot:shard} with different federation sizes and assumptions about the adversary. $K = 100$. Left: Semi-honest protocol with less-dropout tolerable configurations ($\eta = -20$, $\delta >= .05$). Center: Semi-honest protocol with more conservative parameter settings. In this case $\eta = -30$, and the highest tested $\gamma$ is $.5$. Right: Malicious protocol configurations.}
    \label{fig:server_time}
\end{figure*}
}
\newcommand\figclient{
\begin{figure*}[!htb]
    \makebox[\textwidth][c]{\includegraphics[width=1.2\textwidth]{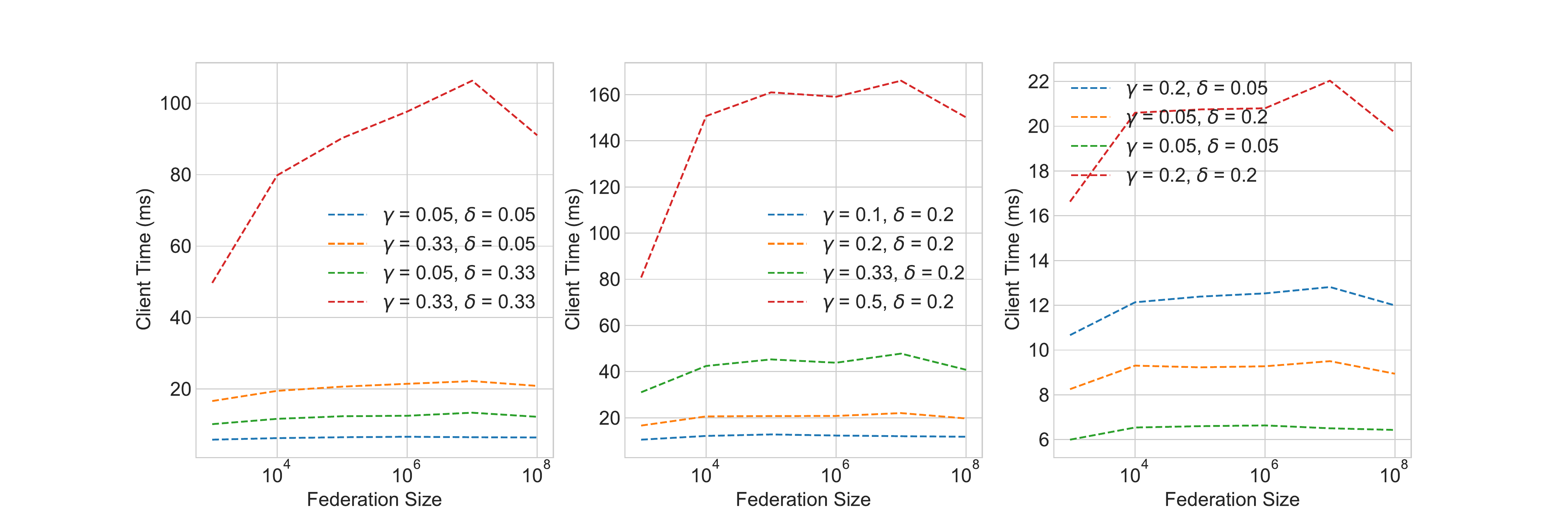}}
    \caption{Per client time for Protocol~\ref{prot:shard} with different federation sizes and assumptions about the adversary.$ K = 100$. Left: Semi-honest protocol with less-dropout tolerable configurations ($\eta = -20$, $\delta >= .05$). Center: Semi-honest protocol with more conservative parameter settings. In this case $\eta = -30$, and the highest tested $\gamma$ is $.5$. Right: Malicious protocol configurations.}
    \label{fig:client_time}
\end{figure*}
}
\newcommand\tabresults{
\begin{table*}
  \centering
    \def\arraystretch{1.2}
    \begin{tabular}{|c|c|c||c|c|c|c|}
    \hline
        Parties & Parties  & Dropouts &  Server &  Client    &  Server  & Client  \\
(Bonawitz) & (\ourprot) & &  (\ourprot) & (\ourprot) & (Bonawitz) & (Bonawitz)\\
    \hline \hline
        500 & 100,000 &    0    &   101   ms &     5424   ms &     2018         ms &   849  ms \\
    \hline
        1000 &  1,000,000 &   0    &   898   ms &     5878   ms &     4887         ms &  1699  ms \\
    \hline
    \hline
        500 &  100,000 &   166  &   101  ms &     5424   ms &     143,389       ms &   849  ms \\
    \hline
        1000 & 1,000,000 &   333   &  898  ms &   5878   ms &     413,767       ms &   1699 ms \\
    \hline
\end{tabular}

    \caption{Comparison to results from Bonawitz et al.~\cite{bonawitz2017aggregation} for 500 and 1000 parties with 0 and 30\% dropout are included for comparison. In order to create these results, we consider our $k = 100$ }
    \label{tab:scale_time}
\end{table*}
}

%% file: intro.tex
\section{Intro}

Efficient \emph{secure aggregation} protocols allow distributed data owners (\emph{clients}) to aggregate secret inputs, revealing only the aggregated output to a (possibly untrusted) server. Secure aggregation protocols can be used to build privacy-preserving distributed systems, including systems for data analytics~\cite{roth2019honeycrisp} and federated machine learning~\cite{kairouz2019advances, kairouz2021distributed}.

The state-of-the-art large vector aggregation protocol~\cite{bonawitz2017aggregation} leverages \emph{masks}---one time pads created with shared random seeds---to encrypt and decrypt the vectors. This reduces communication among parties substantially. Bell et al.~\cite{bell_paper} further reduce communication cost by circumventing the need for a complete communication graph. Rather than sharing a random seed with every other party, each party shares merely with $O(\log{n})$ neighbors.

However, masking-based protocols incur significant communications overhead for short vectors. For a vector of size 100, the Bonawitz protocol results in an \emph{expansion factor} equal to the number of neighbors per party. Expansion factor measures the client communication cost relative to the size of their private inputs. Such a large expansion factor implies that masking protocols provide little to no benefit over the na\"ive solution with small vectors. In the case of dropouts, both protocols undergo a costly unmasking procedure that takes several minutes of server computation time.

In this paper, we propose \ourprot, a highly scalable secure aggregation protocol with dropout robustness. \ourprot is the first sublinear communication complexity protocol to handle dropouts without a recovery communication phase. Table~\ref{tab:complex} presents the computation and communication complexity of \ourprot along with those of the current state-of-the-art for large federation secure aggregation.

\begin{table*}[t]
    \begin{center}
    \begin{tabular}{|c||c|c|c|}
        \hline
      \textit{Setting} & \textit{Bonawitz et al.~\cite{bonawitz2017aggregation}} & \textit{Bell et al.~\cite{bell_paper}} & \textit{\ourprot (ours)} \\ 
        \hline \hline
        Client Communication & $O(n + l)$ & $O(\log n + l)$ & $O(l\log n)$ \\
        \hline
        Client Computation & $O(n^2 + nl)$ & $O(\log^2n + l\log n)$ & $O(l\log^2n)$ \\
        \hline
        Server Communication & $O(n^2+ nl)$ & $O(n\log n + nl)$ & $O(ln)$\\
        \hline
        Server Computation & $O(ln^2)$ & $O(n\log^2n + nl\log n)$ & $O(ln)$ \\
        \hline
    \end{tabular}
    \end{center}
    \caption{Communication and computation complexities of \ourprot compared with the state of the art, for $n$ parties aggregating vectors of size $l$.}
    \label{tab:complex}
\end{table*}

We start with a natural approach to reducing communication complexity: $n$ clients organize into groups of size $O(\log{n})$, aggregate within their groups, and reveal the group's sum to the server. Unfortunately, this approach reveals each group's sum to the server, and the sum of inputs within a small group reveals much more information than the total sum over all $n$ clients.

Our approach addresses this problem via \emph{sharding}. Sharding is a technique borrowed from distributed databases~\cite{CorbettShard, GlendenningShard, MegastoreShard} and scalable blockchains~\cite{LuuShard} where a piece of information is fragmented into pieces (called \emph{shards}) to enhance a desired property (in our case, security). In our \ourprot protocol, each client splits their input into $m \geq 2$ shards, such that each shard in isolation reveals nothing about the input. For shard number $i$, the clients organize into groups of size $O(\log{n})$, sum their $i$th shards using a simple secure aggregation protocol, and reveal the group's $i$th shard sum to the server. The key insight of \ourprot is that the sum of a group's $i$th shard reveals nothing about the sum of the original inputs, as long as \emph{different groups are used for each shard}.

For $m$ shards, \ourprot requires each client to participate in $m$ instances of a simple secure aggregation protocol with only $O(\log{n})$ other clients, matching the communication complexity of the state-of-the-art protocol~\cite{bell_paper}. In most cases, $m=2$ provides sufficient security. Because it is based on threshold secret sharing, \ourprot is robust to dropouts modulo a minimal threshold for construction of the output.

In addition to complexity analysis, our formal results include malicious security of \ourprot in a real-ideal model. We have also implemented \ourprot and performed an empirical evaluation of its performance, demonstrating concrete efficiency of our approach: the computation time for both client and server are less than 100ms, even for federations of size 100 million. \ourprot also provides a significant improvement in concrete communications cost compared to Bell et al.~\cite{bell_paper}, as measured by expansion factor---especially for small private inputs. Moreover, in the presence of dropouts, our approach provides orders-of-magnitude improvement in performance over previous work.

\subsection{Contributions}

\noindent In summary, we make the following contributions:

\begin{enumerate}[leftmargin=20pt]
\item We propose a novel scalable secure aggregation protocol, based on layered secret sharing, with improved concrete computation and communications cost compared to previous work (including an orders-of-magnitude improvement in the presences of dropouts).
\item We prove malicious security of $\ourprot$ in the real-ideal model with modifications both to reflect dropout resistance and to support messaging efficiency in large network settings.
\item We implement our approach and conduct an experimental evaluation demonstrating its concrete efficiency.
\end{enumerate}

%% file: background.tex
\section{Background and Related Work}

\subsection{Secure Aggregation}

\emph{Secure aggregation} protocols are secure multiparty computation
(MPC)~\cite{evans2017pragmatic} protocols that allow a set of clients
to work with a central server to aggregate their secret inputs,
revealing only the final aggregated result. Secure aggregation
protocols have been developed that are robust against both a corrupt
central server and some fraction of corrupt clients, in both the
semi-honest and malicious settings.

The first scalable (1000 parties or more) secure aggregation protocol
is due to Bonawitz et al.~\cite{bonawitz2017aggregation}. In the
Bonawitz protocol, each party generates a \emph{mask} to obscure their
input, and submits the masked input to the server. The clients then
perform pairwise aggregation of their masks, and send the final
aggregated masks to the server. Finally, the server uses the
aggregated masks to reveal the sum of the inputs. The primary
communication cost in this protocol comes from the pairwise
aggregation of masks, which is linear in the number of participating
clients.

Bell et al.~\cite{bell_paper} improve the communication cost of the Bonawitz
approach by layering an additional protocol on top of it. The Bell
protocol prunes the communication graph of the Bonawitz protocol such that
each of the $n$ clients communicates with $\log{n}$ other clients, and
runs the Bonawitz protocol using this graph---reducing communication
cost to be logarithmic in the number of clients.
A complete comparison of asymptotic costs appears in
Table~\ref{tab:complex}, for both existing protocols and our new
approach.

\paragraph{Our Contribution.}
Our novel protocol improves on previous work in three primary ways:
(1) we achieve similar asymptotic complexity to Bell et
al.~\cite{bell_paper} for the client, and improved complexity for the
server; (2) our approach has significantly better \emph{concrete}
communications and computation compared to previous work; (2) our
approach is \emph{orders-of-magnitude} faster than previous work at
handling dropouts during aggregation.

\subsection{Secret Sharing}

Our approach makes extensive use of \emph{threshold secret sharing}. A
$(t, n)$-secret sharing scheme splits a secret into $n$ \emph{shares}
such that at least $t$ shares are required to reconstruct the secret.
Our approach requires a threshold secret sharing scheme with the
following properties:
\begin{itemize}[leftmargin=12pt, itemsep=5pt]
\item $\texttt{share}(t, n, s)$: breaks secret $s$ into $n$ secret
  shares that can reconstruct $s$ with any subset of at least $t$
  shares.
\item $\texttt{reconstruct}$: accepts a set of secret shares $[s]$ as
  input and attempts to reconstruct secret $s$.
\item $\forall a, b: [a] + [b] = [a + b]$ (additive homomorphism).
\end{itemize}
We use Shamir's secret sharing scheme~\cite{shamir1979share}, which
satisfies the above requirements. Our implementation uses
\emph{packed} Shamir secret sharing~\cite{franklinyung}, also known
as batched secret sharing~\cite{baron2015communication}, which speeds up
sharing more than one value at a time.

As we will prove in Section~\ref{sec:protocol-privacy}, the security of 
\ourprot is based on the security guarantee of our secret sharing scheme.
If the secret sharing scheme is secure in the malicious setting, so is
\ourprot. We use a reconstruction scheme similar to Benaloh's~\cite{benaloh86}
to ensure security in the malicious model.

\subsection{Hypergeometric distribution}

The hypergeometric distribution models the process of sampling objects
from a population without replacement. $HyperGeom(t, n, m, k)$ is the 
probability of drawing $t$ successes out of $k$ draws from a population 
of size $n$ which contains $m$ successes. We use the hypergeometric 
distribution to model the probability that a subset of our federation
will or will not be secure and correct.

\subsection{Applications of Secure Aggregation}

The target application for the secure aggregation protocol of Bonawitz
et al.~\cite{bonawitz2017aggregation} was \emph{federated
  learning}~\cite{kairouz2019advances}, a distributed approach to
machine learning. Secure aggregation is particularly useful as a
component in systems for \emph{privacy-preserving deep learning}, in
which clients use their sensitive data to locally compute updates for
a centralized model. A single client's update may reveal that client's
sensitive data, but secure aggregation protocols can be used to
aggregate the updates for learning without revealing any single
client's information. In this context, secure aggregation protocols
operate on gradients or model updates represented by large vectors
(containing hundreds of thousands to hundreds of millions of elements).

To prevent even the information leakage of aggregated updates, secure
aggregation has been combined with \emph{differential
  privacy}~\cite{dwork2014algorithmic} to enable differentially
private federated learning~\cite{kairouz2021distributed,
  truex2019hybrid}. Differential privacy requires the addition of
random noise to ensure privacy; when the central server is trusted,
then the server can be responsible for adding the noise. In our
setting of a potentially untrusted server, each of the clients can add
enough noise that the aggregated results satisfy differential privacy
(as described by Kairouz et al.~\cite{kairouz2021distributed}). The
combination of scalable secure aggregation protocols with differential
privacy allows for a stronger privacy guarantee than either technique
by itself.

Outside of federated learning, the values being aggregated are
typically smaller. Differentially private analytics systems like
Honeycrisp~\cite{roth2019honeycrisp}, Orchard~\cite{roth2020orchard},
and Crypt$\epsilon$~\cite{roy2020crypt} use specialized protocols for
lower-dimensional data in order to scale to millions of participants,
and generally require some trust in the server. Our \ourprot protocol
has the potential to replace these specialized approaches and provide
a stronger threat model, due to its ability to scale to hundreds of
millions of clients.

\subsection{MPC for Machine Learning}

A plethora of MPC protocols have been proposed to accomplish efficient
federated learning. Many of these protocols are designed in a different
threat model than \ourprot. Several take advantage of a semi-honest 
server~\cite{truex2019hybrid}, or use two non-colluding
servers~\cite{ryffel2020ariann, davidson2021star, jayaraman2021revisiting}.
Secure aggregation protocols~\cite{bell_paper, bonawitz2017aggregation}
also leverage MPC techniques, and can be applied to federated learning.
Applications of MPC for federated learning tend to use smaller
federations than what is described in this work~\cite{byrd2020differentially,
xu2019hybridalpha, li2021privacy}.

\subsection{Generic MPC}

MPC protocols can implement any function through arithmetic or boolean
circuits~\cite{yao1986generate, bgw, gmw, bmr, spdz}. These generic MPC
protocols work well in the two-party setting, in semi-honest and malicious
settings, and tend to be optimized for circuit depth. While some of these
protocols can extend to handling hundreds of users, they require a fully
connected communication graph and do not scale to the large federations
studied in this work.

%% file: protocol.tex
\section{Protocol definition}

\begin{figure*}
  \centering
  \includegraphics[width=\textwidth]{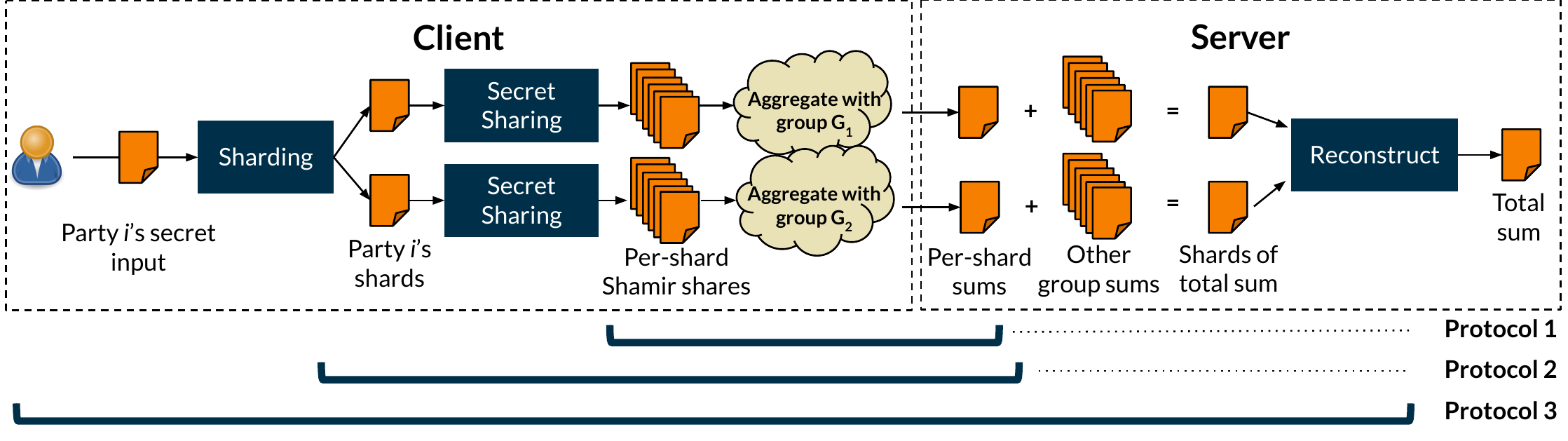}
  \caption{Overview of \ourprot. Each client splits their input into shards, then aggregates each shard in a small group and reveals the result to the server. The server can reconstruct the total sum, but not the sum of any small group's inputs.}
  \label{fig:overview}
\end{figure*}

This section describes $\ourprot$, our novel secure aggregation protocol to emulate Functionality~\ref{prot:ideal}. The ideal functionality sums together the vectors that the trusted third party receives from each client. The output is one vector the same shape as any of the vectors received from any client.

\subsection{Overview}


We implement \ourprot by applying the intuition of sharding to a secret sharing context. 
Sharding, when used in distributed databases or blockchains, refers to breaking information into pieces (called shards) and distributing them among a federation for the sake of security or performance.

In our protocol, we utilize Shamir sharing to break a parties' secret input into shards. 
Those shards are then further fragmented by another round of secret sharing.

A visual overview of \ourprot appears in Figure~\ref{fig:overview}.
The intuition is to secret share each share, and aggregate the secondary shares in small ($O(\log n)$-sized) groups. 
By doing so, we allow parties to aggregate their secrets among small subsets of the federation. 
Their secrets are protected by the redundancy of the multi-level Shamir sharing approach. 
If a small group happens to be controlled by the adversary, the adversary has the ability to learn a share of the secret of each honest party in that small group.
Given the definition security properties of secret sharing, an individual shard is useless on it's own, and the adversary needs to control several specific groups in order to find enough shards to reconstruct an honest party's secret. 

By choosing the number of members in each small group as well as the number of shards into which each secret is broken, we are able to effectively bound the probability of an adversary attacking this protocol in the semi-honest and malicious settings. 

\paragraph{Protocol Overview.}
Protocols~\ref{prot:gagg}, \ref{prot:sub_agg}, and \ref{prot:shard} describe our aggregation method in detail. The three sub-protocols function together as follows:
Protocol~\ref{prot:gagg} describes a simple Shamir sharing based aggregation protocol. Each member of a group sends a share of their secrets to every other member of that group. The parties add their shares and reconstruct the sum of their secrets. This is a well documented extant protocol that we use as a subroutine for sharding.

Protocol~\ref{prot:sub_agg} refers to the process of secure aggregation with subsets of the federation. Where parties Protocol~\ref{prot:gagg} send secret shares to every other party in their federation, the federation in Protocol~\ref{prot:sub_agg} is broken up into a number of smaller groups and each group performs and instance of Protocol~\ref{prot:gagg}. The returned sums from all instances are then added together to calculate the sum of all secret inputs. 

This protocol can aggregate among large federations without revealing private inputs provided that the group size and threshold are selected properly. Our formula for calculating both of those parameters is included in Section~\ref{sec:prob}.

\subsection{Threat Model}
\label{sec:threat}

We adopt the threat model of Bell et al.~\cite{bell_paper}, since it is well-suited to the setting of large federations. Our setting involves two classes of parties: (1) a single \emph{server}, and (2) $n$ \emph{clients}. We assume that the adversary may control \textbf{both} the server \textbf{and} a fraction ($\gamma$) of the clients. $\gamma = \frac{1}{2}$ corresponds to assuming an honest majority of clients; for very large federations, it may be reasonable to assume a smaller $\gamma$. Our use of $\gamma$ is similar to a $(t, n)$-Shamir sharing scheme's security against a $t/n$-sized proportion of clients. Our guarantees have several other parameters, described below (and summarized in Section~\ref{sec:prob}, Table~\ref{tab:vars}).

\paragraph{Semi-honest security (confidentiality).}
In the semi-honest setting, we assume that the server and all clients execute the protocol correctly, but that the adversary-controlled parties (including the server) will attempt to learn the inputs of individual honest clients by observing the protocol's execution. \ourprot guarantees that with probability $1-2^{-\sigma}-2^{-\eta}$, an adversary who controls fewer than $\gamma n$ clients does not learn the input of any honest client.

\paragraph{Malicious security (confidentiality).}
In the malicious setting, we assume that adversary-controlled parties (including the server) may deviate arbitrarily from the protocol. In the malicious setting, \ourprot guarantees that with probability $1-2^{-\sigma}-2^{-\eta}$, an adversary who controls fewer than $\gamma n$ clients does not learn the input of any honest client (i.e. the same confidentiality guarantee as in the semi-honest setting). We prove malicious security in Section~\ref{sec:protocol-privacy}.

\paragraph{Dropouts, correctness, and availability.}
\ourprot separately guarantees availability of the output against $\delta f$ clients dropping out. This guarantee is more important among very large federations because the probability of some dropouts increases as the federation size increases. \ourprot cannot guarantee correctness or availability of the output when the server is malicious. In the event that a malicious server forces parties to dropout, we cannot guarantee availability or correctness, but can guarantee confidentiality of honest inputs. Like Bonawitz et al.~\cite{bonawitz2017aggregation}, and Bell et al.~\cite{bell_paper} we make the assumption that clients are authentic and not simulated for the sake of a Sybil attack. We assume the list of clients is public prior to commencing the protocol, and the existence of secure channels among the parties. As described in previous work~\cite{bonawitz2017aggregation, bell_paper}, this problem can be solved using a Public Key Infrastructure (PKI) or by assuming the server behaves honestly in the initialization round.

\paragraph{Failure probability.}
Traditional MPC security guarantees ensure that there is no chance of an adversary breaking the confidentiality or integrity of a protocol, provided that that adversary is not too strong. In the context of secret sharing, these guarantees inherently limit communication efficiency. For a $(t, n)$- secret sharing scheme, guaranteeing that no adversary smaller than $t$ can compromise security requires that each party communicates with at least $t$ other parties. 

In order to improve communication efficiency, Bell et al.~\cite{bell_paper} and \ourprot specify our security guarantees with small probabilities of failure, which are parameterized by $\sigma$ and $\eta$. $2^{-\sigma}$ is the probability that the security guarantee is not realized, and $2^{-\eta}$ is the probability that the availability guarantee is not realized.

We set $\sigma$ and $\eta$ identically to Bell et al.~\cite{bell_paper} and choose $\sigma = 40$ and $\eta \geq 20$. This relaxation allows \ourprot to significantly reduce communication complexity in exchange for a one-in-a-trillion chance that an adversary can expose private inputs.

\paragraph{Realism of the threat model.}
In real-world deployments (e.g. federated learning or statistical analysis), the server operator generally has a strong incentive to produce correct outputs---obtaining this output is typically the purpose of deploying the system in the first place. Clients, on the other hand, typically care primarily about confidentiality---the final output is being computed for the benefit of the server operator, and its correctness does not benefit the client directly.

Like previous secure aggregation protocols~\cite{bonawitz2017aggregation, bell_paper}, our threat model is designed to align with these incentives. Our primary goal is providing confidentiality for clients; \ourprot does not ensure correctness or availability of the final output when the server is malicious, but the server operator has no incentive to corrupt their own final result.

\paragraph{Comparison of the threat model with related work.}
Compared to the closest related work---the protocol of Bell et al.~\cite{bell_paper}---our threat model is slightly stronger. Our threat model matches that of Bonawitz et al.~\cite{bonawitz2017aggregation} exactly.
Bell et al.~\cite{bell_paper} uses $\alpha \in (0,1]$ to describe the amount of information leaked by a given secure aggregation protocol. For a $n$ party federation, $\alpha$ implies that any party's information will be securely aggregated with at least $\alpha n$ participants. In the protocol of Bell et al., reducing $\alpha$ can improve performance.

The ideal functionality has $\alpha = 1 - \delta - \gamma$. This implies that all honest parties will have their values aggregated together. This is the best we can hope for because parties who drop out might not have input, and malicious parties can subtract their inputs from the ideal functionality's output to obtain the sum of just the honest party's inputs.

\ourprot always ensures the optimal value of $\alpha$. The earlier protocol of Bonawitz et al.~\cite{bonawitz2017aggregation} also ensures the optimal value of $\alpha$, via communication between all pairs of parties.

\begin{algorithm}
  \newcommand{\nonl}{\renewcommand{\nl}{\let\nl\oldnl}}
    \SetAlgorithmName{Functionality}{}{}
    \SetKwInOut{Input}{Input}
    \SetKwInOut{Output}{Output}

    \Input{A set of private vector inputs $s_0 \dots s_n$.}
    \Output{The sum of all values $s_0 \dots s_g$, which we denote as $s$.}
    {\nonl \textbf{Round 1:} Each party $j$:}
    \begin{enumerate}
        \item send $s_j$ to the trusted third party
    \end{enumerate}
    {\nonl \textbf{Round 2:} Trusted third party}
        $$s \leftarrow \sum_{i=1}^{n} s_i$$.
    \caption{Ideal Functionality}
    \label{prot:ideal}
\end{algorithm}

\setcounter{algocf}{0}

\begin{algorithm}
  \newcommand{\nonl}{\renewcommand{\nl}{\let\nl\oldnl}}
    \SetAlgorithmName{Protocol}{}{}
    \SetKwInOut{Input}{Input}
    \SetKwInOut{Output}{Output}

    \Input{a group of $g$ participants, an input for each participant $s_i$, a threshold $t$}
    \Output{The sum of all values $s_0 \dots s_g$}
    {\nonl \textbf{Round 1:} Each party $j$:}
    \begin{enumerate}
        \item $sh_j^0 \dots sh_j^n \leftarrow \texttt{share}(t, g, s_j)$
        \item sends $sh_j^i$ to party $i \forall i \in [0, g]$.
    \end{enumerate}
    {\nonl \textbf{Round 2:} Each party $j$:}
    \begin{enumerate}
        \item receives $sh_0^j \dots sh_g^j$
        \item $sum_j \leftarrow \Sigma sh_0^j \dots sh_g^j$.
        \item broadcasts $sum_j$.
    \end{enumerate}
    {\nonl \textbf{Round 3:} Each party $j$:}
    \begin{enumerate}
        \item receives $sum_0 \dots sum_g$.
        \item $sum \leftarrow \texttt{reconstruct}(sum_0 \dots sum_g)$.
    \end{enumerate}
    \caption{\texttt{group\_agg}}
    \label{prot:gagg}
\end{algorithm}

\begin{algorithm}
  \newcommand{\nonl}{\renewcommand{\nl}{\let\nl\oldnl}}
    \SetAlgorithmName{Protocol}{}{}
    \SetKwInOut{Input}{Input}
    \SetKwInOut{Output}{Output}

    \Input{a partition $P$ of $n$ participants, a group size $g$, a threshold $t < g$, each participant supplies their secret input $s_i$} 
    \Output{The sum of all values $s_0 \dots s_n$ which we call $S$}
    {\nonl \textbf{Round 1:} Each party $j$:}
    \begin{enumerate}
        \item partitions $P$ into groups of size $g$. Groups are partitioned deterministically such that each party creates the same set of groups. Party $j$ is a member of one group: $G_j$. See section~\ref{sec:group} for more information.
        \item $sum_j \leftarrow $\texttt{group\_agg}$(G_j, s_j)$
        \item sends $sum_j$ to the server.
    \end{enumerate}
    {\nonl \textbf{Round 2:} The server:}
    \begin{enumerate}
        \item receives $sum_0 \dots sum_n$
        \item verifies that $sum_i = sum_h$ if parties $i, h$ are in the same group. If this is not true for all groups, $ABORT$.
        \item $S \leftarrow \Sigma_{i = 0}^n sum_i /g$
    \end{enumerate}
    \caption{\texttt{sub\_agg}}
    \label{prot:sub_agg}
\end{algorithm}

\begin{algorithm}
  \newcommand{\nonl}{\renewcommand{\nl}{\let\nl\oldnl}}
    \SetAlgorithmName{Protocol}{}{}
    \SetKwInOut{Input}{Input}
    \SetKwInOut{Output}{Output}

    \Input{Set of $n$ participants $P$ where each participant $i$ has a value $v_i$, a group size $g$ and a number of shards $m$ (almost always $2$). A threshold $t < g$. }
    \Output{The sum of all values $v_0 \dots v_n$ which we denote $V$.}
    {\nonl \textbf{Round 1:} Each party $j$:}
    \begin{enumerate}
        \item shards $sh_0^j \dots sh_m^j \leftarrow \texttt{share}(m, m, v_j)$
        \end{enumerate}
     
    {\nonl \textbf{Round 2:} All parties} \\
        \For {$ i \in \{0 \dots m\}$} {
                \begin{enumerate}
                \item parties agree on $Perm$, a permutation  \\ of the participant list.
                \item $sum_i \leftarrow $\\ \texttt{sub\_agg}$(Perm,\ g,\ t,\ sh_i^0 \dots sh_i^n)$
                \end{enumerate}
            } 

    {\nonl \textbf{Round 2:} The Server:}
    \begin{enumerate}
        \item $V \leftarrow \texttt{reconstruct}(sum_0 \dots sum_m)$.
    \end{enumerate}
    \caption{\ourprot}
    \label{prot:shard}
\end{algorithm}

\subsection{Example Protocol Trace}

The following small example illustrates \ourprot in action and
higlights its features. Suppose we have parties $A$, $B$, $C$, $D$ with
secrets in $\mathbb{F}_2$. First, each party breaks their secret into
shards as shown in the table below. For the sake of this example,
parties use additive secret sharing for shard generation.
$$\begin{tabular}{|c||l|l|}
\hline
    Party & Secret & Shards \\
\hline
\hline
    A & 1 &  1, 0 \\
\hline
    B & 1 &  0, 1 \\
\hline
    C & 0 &  0, 0 \\
\hline
    D & 0 &  1, 1 \\
\hline
\end{tabular}$$
The parties will now perform the \texttt{sub\_agg} protocol on
their two shards. This includes a partitioning of parties into 
subsets.
$$\begin{tabular}{|c||c|c|}
    \hline
    Round & subset 1 & subset 2 \\
    \hline
    \hline
    1 & \{A, B\} & \{C, D\} \\
    \hline
    2 & \{D, B\} & \{C, A\} \\
    \hline
\end{tabular}$$
We note that for groups of size 2, it is trivial for an adversarial
party to determine their group mate's shard in both rounds. That said,
the mechanism of sharding, together with partitioning, prevents the
adversary from learning the other shard, thus maintaining the privacy
of inputs. In this example, if $B$ is an adversary, it can learn $A$'s
first shard and $D$'s second shard. However, it cannot determine $A$
or $D$'s other shards either directly- due to choice of partitions- or
indirectly- because it knows nothing about $C$'s shards.  This
outlines the importance of proper group selection to ensure protocol
security. If we used the same groups for rounds 1 and 2, then $B$
would learn $A$'s secret, etc. Of course, if two parties $B$ and $C$
are corrupt, then they may collude to obtain the secrets of $A$ and
$D$, but we assume an honest majority.

Once the parties are broken into groups, they perform
\texttt{group\_agg} and aggregate their to find the sums of each
sharding round. For the sake of brevity we consider 
\texttt{group\_agg} a black box that returns the sum of shards.
$$\begin{tabular}{|l||l|}
    \hline
    Protocol & Result \\
    \hline
    \hline
    \texttt{group\_agg}(\{A, B\}) & 1 \\
    \hline
    \texttt{group\_agg}(\{C, D\}) & 1 \\
    \hline
    \texttt{sub\_agg}(Round 1) & 0 \\
    \hline
    \hline
    \texttt{group\_agg}(\{D, B\}) & 0 \\
    \hline
    \texttt{group\_agg}(\{A, C\}) & 0 \\
    \hline
    \texttt{sub\_agg}(Round 2) & 0 \\
    \hline
\end{tabular}$$
In all cases \texttt{group\_agg} returns the sum of the shards applied
as input. In round 1, we have $1 + 0 = 1$ and $0 + 1 = 1$ for the
shards of $A$, $B$, $C$, and $D$ respectively. These group-level sums
are aggregated per the \texttt{sub\_agg} protocol to obtain the sum of
0. An identical process is applied to the round 2 shards to calculate
their sum, which is also $0$.

The final step of \ourprot is to reconstruct the output $V$ from
the sharding round sums. Because we are using additive secret
sharing in this example, this process is simply:
$$V = \texttt{sub\_agg}(Round 1) + \texttt{sub\_agg}(Round 2) = 0 + 0 = 0$$
This is correct- the sum of all inputs is $1 + 1 + 0 + 0 = 0$ in $\mathbb{F}_2$.

\subsection{Protocol Privacy}
\label{sec:protocol-privacy}

\subsubsection{Threat models}

We consider a semi-honest, and a malicious secure threat model parameterized by $\gamma$ and $\delta$ as described in section~\ref{sec:threat}. With respect to protocol execution, the semi-honest and malicious models are differentiated by the security of the secret sharing scheme. If \ourprot is implemented with semi-honest secure secret sharing, then \ourprot is secure in the semi-honest model. If \ourprot is implemented with malicious secure secret sharing, then \ourprot is secure in the malicious model. 

Because the semi-honest threat model is a specific case of the malicious threat model, we prove security in the malicious model. In the malicious model, we expect arbitrary deviations from the protocol from both malicious clients and the server. Furthermore, we expect the server and malicious clients to collaborate.

We do, however, assume that the server is not simulating parties as part of a Sybil attack. Preventing this behavior can be solved with public key infrastructure, and we consider protection against this type of attack out of scope for \ourprot. This is the only restriction we apply to server behavior for the sake of input confidentiality. It is also worth noting \ourprot ensures correctness and availability against a $\delta$ fraction of clients dropping out, but does not guarantee correctness or availability against a dropped out server.

\subsubsection{Malicious Security}

Suppose the ideal functionality of addition as $F$, an adversary $A$. Let $v_i$ and $x_i$ be input and view of client $i$ respectively.  Let $V$ be the output of $\pi$.

Let $U$ be the set of clients. Let $C \subset U \cup  \{S\}$ be the set of corrupt parties, and $D \subset U$ be the set of dropped out parties. The set of honest parties is  $H = U \setminus (C \cup D)$.

In this proof, we consider the dropped out parties as a part of the adversary without loss of generality.

\begin{theorem}

     There exists a PPT simulator \texttt{SIM} such that for all $U$, $|C| \leq \gamma |U|$, and $|D| \leq \delta |U|$ 

\[ \texttt{REAL}_{\pi, A}(n; x_{H}) \equiv \texttt{IDEAL}_{F, \texttt{SIM}}(n, x_{H}) \]

\end{theorem}

The intuition behind this statement is that no such adversary can exist on our protocol that is more powerful than an adversary against the ideal functionality. 
 
\begin{proof}
Proven through the hybrid argument. We assume that any honest party will $ABORT$ if they receive an ill-formed message, an untimely message, or an abort from any other party. Furthermore, we assume secure channels between each pair of parties. 

\begin{enumerate}
    \item This hybrid is a random variable distributed exactly like $\texttt{REAL}_{\pi, A}(n; x_{H})$. 
    \item In this hybrid $\texttt{SIM}$ has access to all $\{x_i | i \in U\}$. $\texttt{SIM}$ runs the full protocol and outputs a view of the adversary from the previous hybrid.
    \item In this hybrid, $\texttt{SIM}$ generates the ideal inputs of the corrupt and dropout parties using a separate simulator $\texttt{SIM}_g$. These sets of inputs, $x_C$ and $x_D$, contain a field element or $\bot$ for each corrupt or dropout party respectively. Through this process, $\texttt{SIM}_g$ may force the output of $F$ to be any field element or $\bot$. Thus $\texttt{SIM}_g$ is able to produce the same protocol outputs that $A$ is able to in $\texttt{REAL}$, so this hybrid is indistinguishable from the previous hybrid.

    \item In this hybrid, $\texttt{SIM}$ replaces $V$, the output of the protocol, with the known output of the ideal function and the aggregation of all the ideal inputs of the corrupt parties. We exclude the inputs of the dropped out parties. This hybrid is indistinguishable from the previous hybrid with probability $2^{-\eta} $ as defined in Section~\ref{sec:prob}, provided that group assignments satisfy property~\ref{prop:con_sec}

    \item In this hybrid $\texttt{SIM}$ replaces the shards of each honest parties with a secret sharing of a random field elements such that the field elements sum to the output of $\texttt{IDEAL}$. This hybrid is indistinguishable from the previous hybrid with probability $2^{-\sigma}$ as defined in Section~\ref{sec:prob}. This is because the adversary should not have access to enough shares to reconstruct any individual party's secret.

\end{enumerate}
\end{proof}

\subsection{Complexity Analysis}

Suppose $n$ clients with $k$ values to send.

    \subsubsection{Client Computation} $O(k\log^2 n)$. The client needs to break $k$ values into $log n$ values. For a Shamir sharing of $m$ shares takes $O(m^2)$. The addition and reconstruction take $O(log n)$ and $O(k log^ n)$ time respectively. 
    \subsubsection{Client Communication}$O(k\log n)$. The client needs to send $O(\log n)$ clients $O(k)$ values each.
    \subsubsection{Server Computation} $O(nk)$. The server needs to add all of the group sums together and reconstruct the shard-level Shamir shares. This includes processing the output of all parties. The shard-level Shamir share is treated as a constant cost because there are always two shard shares.
    \subsubsection{Server Communication} $O(nk)$. The server receives output from all parties.

\section{Group Assignments}\label{sec:group}

Beyond assigning groups such that they are unlikely to be corrupted and that they are unlikely to dropout, we also would like to assign groups over the two rounds such that the outputs of multiple groups cannot be combined to leak additional information. In particular, Protocol~\ref{prot:sub_agg} exposes the sums of each subgroup. Protocol~\ref{prot:shard} can also release sums of small sets of parties if groups are not chosen carefully.

\paragraph{Information Leakage from Overlapping Groups.}

For simplicity we set $m = 2$, which is also consistent with our evaluation. However, we conjecture that the results in this section are easily generalized to $m > 1$. Let $R_1$ and $R_2$ be the sets of groups used in the two respective invocations of \texttt{sub\_agg} within the for loop of Round 2 for protocol 3. Each party is a member of a group in $R_1$ and a member of a group in $R_2$. They aggregate their first shard with the group in $R_1$ and their second shard with the group in $R_2$.

Consider the case where a single group $G$ is used in both rounds: $G \in R_1 \land G \in R_2$. An adversary can reconstruct the sum of inputs of parties in $G$ by using \texttt{reconstruct} on the outputs of $G$ in rounds 1 and 2. 

Requiring $R_1 \cap R_2 =\{\} $ is not sufficient to prevent such an attack. Suppose there exist some groups $G_1, G_2, G_3, G_4$ such that  $G_1, G_2 \in R_1$, $G_3, G_4 \in R_2$ and $G_1 \cup G_2 = G_3 \cup G_4$. An adversary can reconstruct the sum of parties in groups $G_1 \cup G_2$ by calling \texttt{reconstruct}  on the sum of $G_1$ and $G_2$'s round 1 outputs and the sum of $G_3$ and $G_4$'s round 2 outputs.

\paragraph{Graph background.}
A graph $G = (V, E)$ where $V$ is the set of nodes and $E$ is the set of
edges such that $(i, j) \in E \iff i \in V \land j \in V\  \land$ there 
is an edge between $i$ and $j$. These are undirected graphs so $(i, j)$ and
$(j, i)$ are equivalent.
We consider a subgraph $SG = (V', E')$ where
$V' \subseteq V$ and $E' \subseteq E \land (i , j) \in E' \implies (i \in V' 
\land j \in V')$.

Finally we consider a disconnected subgraph $DG = (V'', E'')$ of $G$ if 
$CC$ is a subgraph of $G$, and $\forall (i,j) \in E,\ i \in V'' \implies
(i, j) \in E''$. In other words, all nodes in a disconnected subgraph of $G$
exclusively have edges to other nodes within the disconnected subgraph. The
disconnected subgraph is \emph{disconnected} from the rest of $G$.

\paragraph{Avoiding Information Leakage.}
In order to ensure that no subset sum can be accessed besides the sum of all honest parties, we require that our honest party communication graph is fully connected. We define a party communication graph as 

\[HG = (V, E) \ s.t.\]
\[V = \{honest\ parties\} \]
\[E = \{ (i,j) |\  \exists\  G \in R_1 \cup R_2 | i \in G \land j \in G\} \]

The honest party communication graph draws connections between any two parties that are in a group together in either round.

\begin{property}\label{prop:conn}

    Suppose a subgraph $SG = (V', E')$ of $HG$.
    $sum('V)$ is recoverable $\implies$ $SG$ is a disconnected subgraph

\end{property}

\begin{proof}
    The proof is by contradiction. Suppose a subgraph $SG = (V', E')$ where the sum of $V'$ is accessible, but $SG$ is not a disconnected subgraph.

    Because $SG$ is not a disconnected subgraph, we know
    $$ \exists i, j \in V | i \in V' \land j \notin V' \land (i, j) \in E$$

    Ultimately, there is an edge between a node in $SG$ and a node outside of $SG$.
    From the existence of this edge we know that  $i$ and $j$  were in a group together for one of the sharding rounds.
    This implies that one of $i$'s shards is aggregated with one of $j$'s shards, and a sum including either of these shards would have to include the other.
    We reach contradiction here because $j$ is not in $SG$, so the sum of $SG$ is unavailable.

\end{proof}

From Property~\ref{prop:conn}, the requirement that $HG$ remain fully connected emerges.

\begin{property}\label{prop:con_sec}
 $HG.isconnected \implies$ no subset sum leakage.
\end{property}

This property follows directly from Property~\ref{prop:conn}. Because there are no connected components of a fully connected graph, no sums smaller than the one released by the ideal functionality are revealed.

\paragraph{Generating Groups.}
There are conceivably many different ways to generate groups for two rounds of sharding to ensure $HG$ remains fully connected, and different instantiations of this protocol might want to use different group generation methods to adapt to network conditions like geo-location. In our implementation we determine group membership based on a single permutation of the network. Suppose $i$ is the index of some party in our permutation, and $g$ is the group size. Party $i$ is a member of group $i // g$ for the first round, and $(i//g + g* i\%g) \ n$ for the second round. The expression for second round moves the $j^{th}$ member of each group $j$ groups forward. This spreads parties around for the second round sufficiently enough to ensure that $HG$ is fully connected.

%% file: probability.tex
\figneighbors
\figbellexpansion
\figexpansion
\figserver
\figclient
\tabresults
\section{Setting Parameters}
\label{sec:prob}

Our security proofs for \ourprot assume that group size, and the reconstruction threshold are selected appropriately to guarantee security.
This section reasons about the appropriate parameters given specifications about the aggregating environment. All parameters involved in determining security are listed in Table~\ref{tab:vars}. 
Above the double line are the parameters comprising our configuration.
The protocol administrator will selected these parameters to suit their needs.
Below the double line are the $g$, $t$ and $k$, the parameters we feed directly to the protocol.

\begin{table}
    \begin{tabular}{|c|p{5cm}|}
        \hline
        Parameter & Description \\
        \hline
        \hline
        $\sigma$ & $1 - 2^{-\sigma}$ is the probability of secure protocol execution\\
        \hline
        $\eta$ & $1 - 2^{-\eta}$ is the probability of correct protocol execution.\\
        \hline
        $\gamma$ & corrupt fraction of federation \\
        \hline
        $\delta$ & fraction of federation that will drop out \\
        \hline
        $f$ & Number of clients in the federation. \\
        \hline
        \hline
        $g$ & the number of clients in each group. \\
        \hline
        $t$  & the reconstruction threshold in each group. \\
        \hline
        $k$  & the number of values to be shared at once. \\
        \hline
    \end{tabular}
    \caption{independent and dependent variables to ensure protocol security}
    \label{tab:vars}
\end{table}

We would like to set $g$ and $t$ to ensure that two events do not happen.
\begin{enumerate}
    \item A group is corrupted (more adversaries than $t$).
    \item A group cannot reconstruct its sum (more dropouts than $g - t - k$).
\end{enumerate}

We can guarantee with absolute certainty that these two events do not happen if we trivially set $g = f$ and $t > f\gamma$ and $g - t - k > f\delta$.
However that clearly leads to poor protocol performance. 
Instead, we use the same convention as Bell et al.~\cite{bell_paper} and select $g$, $t$, and $k$ to keep the probability of events (1) and (2) very low.
This is reflected in our parameters in the way of $\sigma$ and $\eta$.
They are defined as $P[(1)] < 2^{-\sigma}$ and $P[(2)] < 2^{-\eta}$.

To determine how likely these events are over the entire federation, we first start by determining their probability at the group level.
 Suppose we have a group of size $g$, and a federation of size $n$. The probability of an individual belonging to a group with $i$ corrupt clients is hypergeometric.  
\[ HyperGeom(i, n - 1, \gamma n, g) \]
We are sampling clients to be corrupt without replacement from a population of size $n - 1$ with $\gamma n$ clients in it because we assume that one client in each group is honest.
In the incredibly unlikely event that a group is entirely comprised of corrupt clients, it is inconsequential to the security of the protocol because no honest client can have their inputs exposed by this event.
Furthermore, they could change the final output of the protocol, but an attack of this variety is no more powerful modifying the adversarial client's inputs to the protocol.

We can use the CDF of the hypergeometric distribution to calculate the probability that one group is not corrupted.
\[ p_nc = HyperGeomCDF(t - 1, n - 1, \gamma n, g) \]
Similarly, the probability of a group reconstructing in spite of dropouts is a CDF of a hypergeometric distribution:
\[ p_nd  = 1 - HyperGeomCDF(g - t - k, n - 1, \delta n, g) \]
Packed secret sharing requires $t + k - 1$ shares to reconstruct a secret.
To use malicious reconstruction, we require $t + k$ shares, so we require that fewer than $g - t - k$ clients in that group dropout.

Finally we need to consider the security and reliability of all groups.
We do so by calculating the probability that all groups are secure and reconstruct properly, then use the complement of these values. 
\[ p_{corrupt}  = 1 -  p_nc^{2n/g} \]
\[ p_{dropout}  = 1 -  p_nd^{2n/g} \]
The exponent $2n/g$ is the total number of groups over both sharding rounds. Finally we take the negative log of our probabilities to compare them to the security parameters $\sigma$ and $\eta$.
\[ \sigma \leq -\log_2(p_{corrupt}) \]
\[ \eta \leq -\log_2(p_{dropout}) \]
These formulas, allow aggregators to specify specific security and correctness parameters $(\sigma, \eta)$, and assume the probability of corrupt or dropped out individuals $(\gamma, \delta)$, and calculate $g$, $t$, $k$.  We implement a search algorithm to determine the minimum number of neighbors each client requires for a given set of security parameters.

%% file: evaluation.tex
\section{Evaluation}

This section evaluates the concrete performance of \ourprot with respect to communication and computation. Through a series of experiments, we will answer the following research questions:

\begin{enumerate}
    \item[\textbf{RQ1}] How does \ourprot scale to large federations?
     \item[\textbf{RQ2}] How does \ourprot handle vector length?
     \item[\textbf{RQ3}] In practice, what are the computational demands of \ourprot?
\end{enumerate}

\paragraph{Implementation.}
For our experiments, we implemented a simulation of \ourprot in Python, using numpy to perform field arithmetic. We implemented packed Shamir secret sharing based on~\cite{dahl_2017}. The code used in our experiments is available as open source on GitHub.\footnote{Redacted for review}

\paragraph{Comparison to Previous Work.}
Our comparisons to the protocols of Bonawitz et al.~\cite{bonawitz2017aggregation} and Bell et al.~\cite{bell_paper} are based on concrete results given in their papers, or calculated based on analytical bounds they give (e.g. for expansion factor and number of required neighbors).

\subsection{Communication Performance}

To answer \textbf{RQ1}, we calculate the communication cost per client for various federation configurations and assumptions. These configurations align closely with those tested by~\cite{bell_paper} in order to provide a clear comparison between our approaches. Federation range from $1000$ to $100,000,000$ parties in these experiments. 

Figures~\ref{fig:neighbors_dp},~\ref{fig:expansion_dp},~\ref{fig:server_time}, and~\ref{fig:client_time} reference semi-honest and malicious threat models. We note that the threat model of \ourprot is dictated by the security of the secret sharing primitive. The primary difference between our semi-honest and malicious secure secret sharing primitives is the malicious secure primitive uses a slower reconstruction technique, and requires one more share per reconstruction.

Each configuration is determined by the parameters described in Table~\ref{tab:vars}: $\sigma$, $\eta$, $\delta$, $\gamma$, $k$, and the federation size. We used a modified binary search to determine the group size and threshold that would appropriately satisfy the constraints formed by the fixed parameters applied to the probability formulas defined in Section~\ref{sec:prob}. Because each party participates in two groups, one for each shard, the total number of neighbors is simply twice the group size.

Figure~\ref{fig:neighbors_dp} displays these results for Protocol~\ref{prot:shard}. We see the expected $O(\log{n})$ trend with respect to the number of neighbors required. Both protocols require a comparable number of neighbors to Bell et al.~\cite{bell_paper}, and substantially fewer shares than the na\"ive approach. Notably, using the malicious protocol has very little effect on the communication complexity. 

To answer \textbf{RQ2}, we evaluate the expansion factor of Protocol~\ref{prot:shard} using packed secret sharing for group level aggregations. A scalable protocol with respect to vector size will have small expansion factors. Expansion factor measures the amount of communication required for a protocol as a multiple of the required communication for the ideal functionality. In our case Expansion factor is:
\[EX = \left( \frac{num\_neighbors}{k} \right) \cdot \log(field\_size)\]
Figure~\ref{fig:expansion_dp} contains the results. The results show that expansion factor depends on the level of robustness against dropouts and malicious clients, but is consistent across federation sizes.

\paragraph{Comparison with Bell et al.~\cite{bell_paper}.}
Figure~\ref{fig:expansion_bell} compares the expansion factor of \ourprot against the protocol of Bell et al.~\cite{bell_paper}.
The amortized number of shares required to represent each value is relatively small considering that we secret share the entire vector. Packing is especially useful in cases where the expected number of dropouts is low.  We calculate the expansion factor for Bell et. al.'s protocol based on the formula in~\cite{bonawitz2017aggregation}, but replacing the federation size with number of neighbors to reflect the optimized communication graph. With small vectors, our protocol provides a substantially smaller expansion factor.  Our protocol's expansion factor remains constant or monotonically decreases as vector size increases. For very large vectors (100k+ elements), prior work~\cite{bell_paper, bonawitz2017aggregation} provides a smaller expansion factor.

\subsection{Computation Performance}

In this section we hope to answer \textbf{RQ3} by simulating \ourprot on large federations and reporting client and server computation performance. We implement our protocol in python and run simulations in a single thread on an AWS z1d.2xlarge instance with 64 Gb of memory~\cite{aws}. Our timing experiments are designed to compare our concrete computation performance with that of prior work. Following the experimental designs of Bonawitz et al.~\cite{bonawitz2017aggregation}, we ignore communication latency and throughput in our experiments.

Figures~\ref{fig:server_time} and~\ref{fig:client_time} respectively plot the server and client computation times for Protocol~\ref{prot:shard}. Even for a federation of $100,000,000$ parties, aggregating $100$ values per client requires less than a second of server computation and less than a 10th second of computation per client for all corruption and dropout assumptions we examined. Provided that dropouts do not increase beyond the assumption made when configuring parameters, the protocol will achieve the correct result with no additional computational cost.  

In order to demonstrate the impact of working with dropouts without additional computation, we partially simulate the use of our protocol to aggregate large vectors and compare with the concrete results presented in Bonawitz et. al.~\cite{bonawitz2017aggregation}. To aggregate a vector with $100,000$ elements, we simply repeat the protocol with $k = 100$ for $1000$ iterations. We simulate a subset of groups and use dummy values for the remainder of the group inputs to server to reduce the total simulation time, and to reduce the effect of memory pressure on client level simulations.

\paragraph{Comparison with Bell et al.~\cite{bell_paper}.}
The work of Bell et. al.~\cite{bell_paper} improves on the number of neighbors required of each client substantially,  we run our protocol on substantially larger federations than the ones considered in ~\cite{bonawitz2017aggregation} to create a more fair comparison to Bell et. al. Our results are included in Table~\ref{tab:scale_time}.

While we see that the sharding approach does not perform as well in the case with no dropouts, adding a few dropouts drastically harms the server computation time of the masking based approach. These amounts of dropouts are substantial in the smaller federations used in ~\cite{bonawitz2017aggregation}, but both prospective numbers of dropouts are far more realistic considering the much larger federations we consider here as they are well less than $1\%$. For this comparison, our approach tolerates $5\%$ dropouts, so we could potentially further increase the number of dropouts at no cost to the sharding approach.

%% file: conclusion_new.tex
\section{Conclusion}\label{sec:con}
We propose a new highly scalable secure aggregation protocol, \emph{\ourprot}, with much better performance compared to prior work \cite{bell_paper} in settings with small vectors or many dropped out parties. 
\ourprot scales gracefully to accommodate hundreds of millions of parties while requiring only hundreds of connections per party in the vast majority of settings. 
Defense against malicious adversaries requires little modification of the protocol, and does not substantially affect communication or computation costs--we simply require one additional share per group, and perform the reconstruction twice. 
Our empirical results show that \ourprot can aggregate over very large federations with a small computational cost. 
Small vector secure aggregation protocols have applications in distributed data analytics as well as smaller machine learning models. Histograms, random forests, logistic regression, and small neural networks would all benefit from protocols enabling short vector aggregation~\cite{fedforests,fedlog}. Thus our technology has potentially broad applications. 
Our experiments suggest that $2$ shards per party is optimal for this
protocol, however tighter approximations of the probability of a
security failure could suggest otherwise.  Furthermore, more rounds of
sharding open the possibility of packed secret sharing within the
sharding round, and a protocol that better supports wider vectors.
Investigation of these threads are future work.